\UseRawInputEncoding
\documentclass[11pt,oneside,reqno]{amsart}
\newtheorem{theorem}{Theorem}[section]
\usepackage[left=2.cm,top=1.5cm,bottom=1.5cm,right=2.0cm]{geometry}
\usepackage{latexsym}
\usepackage{amssymb,amsmath,amsfonts}
\usepackage[pagewise]{lineno}
\usepackage{multirow}
\usepackage{graphicx}
\usepackage{cite}
\theoremstyle{definition}
\newtheorem{definition}[theorem]{Definition}

\newtheorem{proposition}[theorem]{Proposition}

\newcommand{\R}{\mathbb R}
\newcommand{\F}{\mathbb F}

\newcommand{\Scal}{\mathcal{S}}

 \global\parskip 6pt \oddsidemargin
0.1cm \evensidemargin 0.4cm \headheight 0.1cm \topmargin -2.0cm
\begin{document}
\pagenumbering{arabic}
\title[A Solow-Swan framework for economic growth]{A Solow-Swan framework for economic growth with memory effect}
\author{M.O. Aibinu$^{1,2,3,6,*}$, K.J. Duffy$^{4,6}$, S. Moyo$^{5,6}$}
\address{$^{1}$Department of Mathematics and Statistics, University of Regina, Regina, SK S4S 0A2, Canada}
\address{$^{2}$Ingram School of Engineering, Texas State University, TX 78666, United States}
\address{$^{3}$School of Computer Science and Applied Mathematics, University of the Witwatersrand, Johannesburg 2050, South Africa}
\address{$^{4}$School of Mathematics, Statistics \& Computer Science, University of KwaZulu-Natal, Durban 4000, South Africa}
\address{$^{5}$Department of Applied Mathematics and School for Data Science and Computational Thinking, Stellenbosch University, Stellenbosch 7602, South Africa}
\address{$^{6}$National Institute for Theoretical and Computational Sciences (NITheCS), South Africa}
\email{$^*$moaibinu@yahoo.com}

\begin{abstract}
The Solow-Swan equation is a cornerstone in the development of modern economic growth theory and continues to attract significant scholarly attention. This study incorporates memory effects into the classical Solow-Swan model by introducing a formulation based on the Caputo fractional derivative. A comparative analysis is conducted between the integer-order and fractional-order versions of the model to examine the influence of fractional dynamics on capital accumulation. The findings reveal that the inclusion of a fractional-order derivative significantly affects the trajectory and long-term stability of capital, offering a more flexible and comprehensive framework for modeling economic growth processes.
\end{abstract}


\maketitle

\section{Introduction}\label{sse6}
 The Solow-Swan model is a cornerstone of modern economic growth theory. It describes the evolution of capital per worker over time and serves as a foundational tool in both theoretical and applied economics. In its classical form, the model is governed by the differential equation
\begin{equation}\label{sse1}
\frac{dk(t)}{dt} = k(t)\left(pk^{\mu-1}(t) - q\right), \quad k(0) = k_0,
\end{equation}
where $k(t)=K(t)/L(t)$ denotes the capital-labour ratio at time $t,$ with $K$ representing capital and 
$L$ labour. The parameters $p, q$ and $\mu$ are positive constants that characterize the production function and capital depreciation. Equation \eqref{sse1} captures the dynamic behaviour of capital accumulation and is extensively covered in undergraduate economics curricula as well as in contemporary economic research (e.g., \cite{Brunner1}). Over the years, the Solow-Swan model has been extended to a wide array of domains, including climate change \cite{AziziJH}, corruption \cite{Elmslie1}, education \cite{Breton}, and resource exploitation \cite{CayssialsP}. These applications demonstrate the flexibility and relevance of the model in capturing long-term economic dynamics across diverse contexts. Several studies have contributed to the enhancement of the analytical and empirical foundations of the model. For instance, Brunner et al. \cite{Brunner1} employed least squares to estimate model parameters using the 2021 Penn World Table \cite{PWT2021}, a well-established source of economic growth data collected annually since 1950 \cite{Feenstra2015}. Khoo et al. \cite{KhooLHB} considered a hybrid Solow-Swan model, leveraging neural ordinary differential equations, a recently proposed machine learning approach. Akhalaya and Shikhman \cite{AkhalayaS} analyzed the stability of spatially homogeneous equilibria, and P$\check{r}$ibylov$\acute{a}$ \cite{Pribylova} introduced a time-varying labour growth rate $q(t)$ to generalize the model. Despite wide application of the model, existing formulations typically rely on classical (integer-order) derivatives, which may not adequately capture complex dynamics inherent in real-world economic systems, particularly those with memory or hereditary characteristics. In contrast, Fractional Differential Equations (FDEs), which incorporate derivatives of non-integer order, are increasingly recognized as powerful tools for modelling processes with memory effects. Owing to their nonlocal nature, fractional derivatives account for the entire history of a system’s evolution, offering a more realistic representation of phenomena where past states influence current behaviour. In economics, memory effects are particularly relevant. Many economic processes, such as investment decisions, consumption patterns, and technological adoption, exhibit long-term dependencies. While FDEs have been successfully applied in various disciplines, their integration into the Solow-Swan growth framework remains relatively unexplored (see, e.g., \cite{TejadoPV}). Notably, Tarasov \cite{Tarasov} emphasized the suitability of fractional models for capturing such effects in economic systems.

This paper aims to fill this gap by incorporating memory effects into the Solow-Swan model through the use of fractional-order derivatives. Specifically, this paper proposes and analyzes a fractional version of equation \eqref{sse1} using the Caputo derivative, which is well-suited for initial value problems due to its treatment of initial conditions in the classical sense. The paper compares the behaviour of the model under integer-order and fractional-order formulations, with particular emphasis on the influence of the fractional order and the scaling parameters $p$ and $q$ on capital dynamics. Through this extension, this study provides new insights into the temporal evolution of capital accumulation in systems where historical influences play a significant role.

\section{Preliminaries}
This section introduces essential definitions and mathematical tools required for the analysis in this paper. These include the Sumudu transform, the Caputo fractional derivative, properties of the Mittag-Leffler function, and elements related to the Adomian decomposition method.
\begin{definition} ({\bf Sumudu Transform})\\
Let $$\F=\left\{k(t): \exists \ Q, \eta_1, \eta_2 >0, |k(t)|<Q\exp{\left(\frac{|t|}{\eta_j}\right)}, \mbox{for} \ t\in (-1)^j\times [0, \infty) \right\},$$
be a set of functions with suitable exponential growth. The {\bf Sumudu Transform (ST)} of a function $k(t)\in \F$ is defined by (see, e.g., \cite{Belgacem2})
\begin{equation}\label{em1}
\Scal [k(t)]=\int^{\infty}_{0}k(tu)e^{-t}dt, \ u\in (-\eta_1, \eta_2).
\end{equation}
The ST is a linear integral transform with several desirable properties, including unit preservation and domain scaling, which make it attractive for solving differential equations \cite{TunG1}. Let $K(u)=\Scal[k(t)].$ Then, the ST of the $nth$-order derivative of $k(t)$ is given by (see, e.g., \cite{Moltot1, Belgacem2, Watugala1, Belgacem1})
\begin{equation}\label{sumud20}
\mathcal{S}[k^{(n)}(t)] = \frac{1}{u^n} \left[K(u) - \sum_{i=0}^{n-1} u^i k^{(i)}(0)\right].
\end{equation}
In particular, the first-order case simplifies to:
\begin{equation}\label{sumud19}
\mathcal{S}[k'(t)] = \frac{1}{u} \left[K(u) - k(0)\right].
\end{equation}
\end{definition}
\begin{definition} ({\bf Caputo Fractional Derivative}) \\
For a function $k(t)$ and order $\alpha\in(0, 1),$ the {\bf Caputo fractional derivative} is defined as
$$^C_aD^{\alpha}k(t)=\frac{1}{\Gamma\left(1-\alpha\right)}\int^{t}_{a}(t-\eta)^{-\alpha}k'(\eta)d\eta, a>0.$$
The Caputo derivative is widely used due to its compatibility with classical initial conditions. The ST of the Caputo derivative with lower terminal $a=0$ is given by (see, e.g., \cite{Bodkhe})
\begin{equation}\label{em5}
\mathcal{S}[,^C_0D^{\alpha}k(t)] = u^{-\alpha} \left[K(u) - k(0)\right].
\end{equation}
\end{definition}

\begin{proposition} ({\bf Convolution Property of the ST})\\
Let $\psi, \zeta)( :[0, \infty)\rightarrow \R.$ The classical {\bf convolution product} is defined by
$$(\psi * \zeta)(t)=\int^{t}_{0}\psi(t-x)\zeta(x)dx.$$
The ST of the convolution satisfies the identity:
\begin{eqnarray*}
\Scal \left[(\psi * \zeta)(t)\right]&=&u.\Scal[\psi(t)].\Scal[\zeta(t)]=u\psi(u)\zeta(u).
\end{eqnarray*}
\end{proposition}

\begin{definition}\label{em20}  ({\bf Mittag-Leffler Function})\\
The {\bf Mittag-Leffler function} is a generalization of the exponential function and is defined by
$$E_{\alpha}(t)=\sum_{n=0}^{\infty}\frac{t^n}{\Gamma\left(\alpha n+1\right)}, \alpha >0.$$
Some important results involving its ST include (see, e.g., \cite{Nanware}):
\begin{itemize}
	\item [(i)]$\Scal\left[E_{\alpha}\left(-at^{\alpha}\right)\right]=\frac{1}{1+au^{\alpha}},$
	\item [(ii)]$\Scal\left[1-E_{\alpha}\left(-at^{\alpha}\right)\right]=\frac{au^{\alpha}}{1+au^{\alpha}}.$
\end{itemize}
\end{definition}

\begin{definition} ({\bf Adomian Polynomials})\\
Let a nonlinear term $N[k]$ be present in a differential equation with solution expressed as a power series:
$$k(t)=\sum_{n=0}^{\infty}k_nt^n, ~~N[k(t)]=\sum_{n=0}^{\infty}A_nt^n.$$ 
The coefficients $A_n$ are known as {\bf Adomian polynomials}, used in the Adomian Decomposition Method (ADM) to handle nonlinearity. They are given by (see \cite{Adomian1, Adomian2}):
$$A_n=\frac{1}{n!}\left[\frac{d^n}{d{x}^n}f\left(\displaystyle\sum_{i=0}^{\infty}{x}^ik_i\right)\right]\bigg|_{x=0}.$$
The first few Adomian polynomials are:
$$\begin{cases}
A_0=f(k_0),\\
A_1=k_1f'(k_0),\\
A_2=k_2f'(k_0)+\frac{k_1^2}{2!}f''(k_0)\\
A_3=k_3f'(k_0)+k_1k_2f''(k_0)+\frac{k_1^3}{3!}f'''(k_0)\\
\vdots
\end{cases}$$
Each polynomial $A_n$ depends on $\left\{k_0,\cdots, k_n\right\},$ enabling a recursive construction of the series solution to nonlinear problems.
\end{definition}
\section{Main Results and Methodology}
The principal contribution of this paper begins with the investigation of the classical Solow-Swan model and its extension via fractional calculus to incorporate memory effects. The paper focuses on obtaining approximate analytical solutions for both the classical and fractional versions of the model using a hybrid approach based on the ST and ADM (see, e.g., \cite{LiuC, Aibinu7, Aibinu6}). The influence of the fractional-order derivative on the dynamics of capital accumulation is also examined.

\subsection{The Solow-Swan Model with Integer-Order Derivative}
\begin{figure}
\centering
\includegraphics[width=9.0cm ,height=6.0cm]{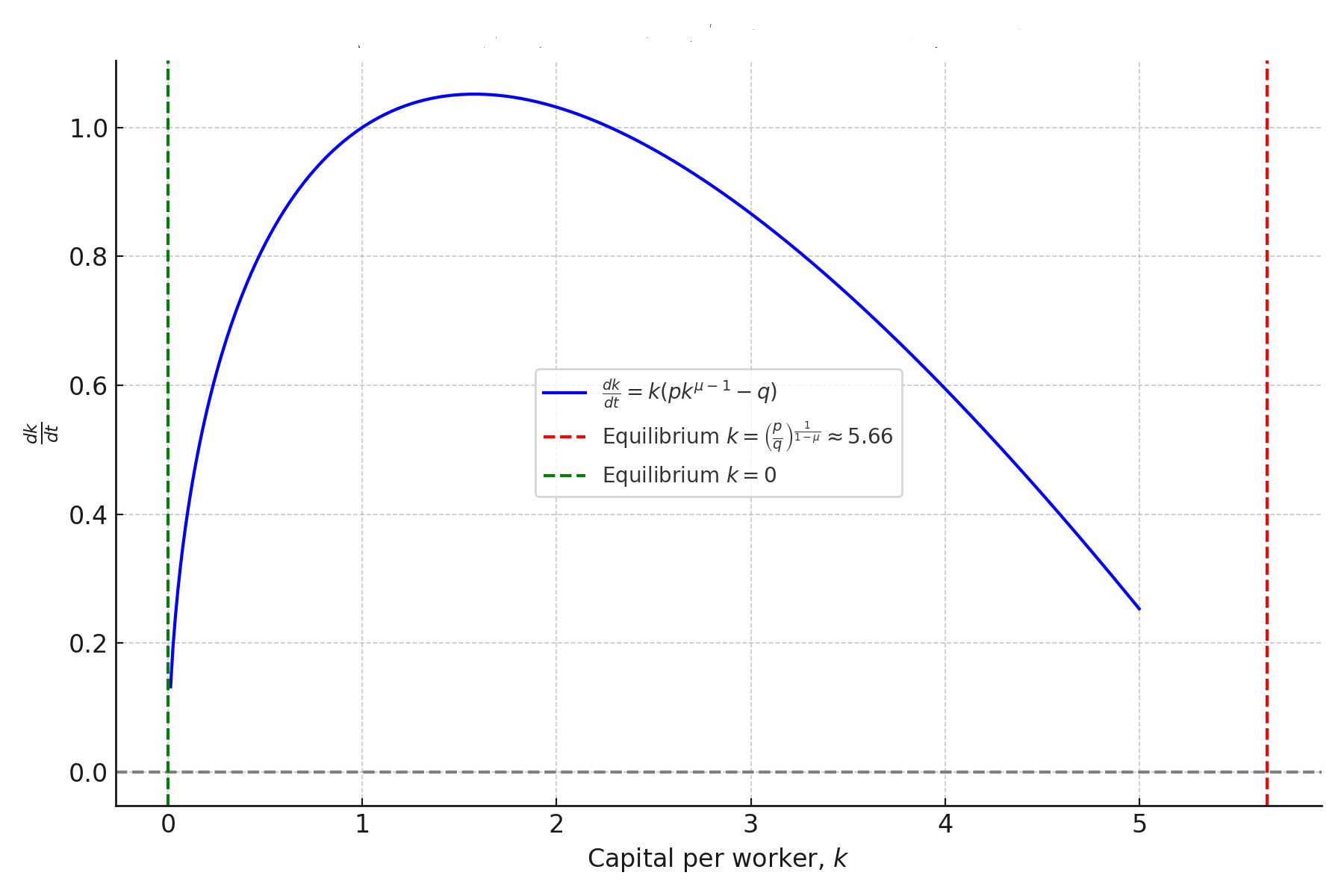}
 \caption{Equilibrium points of the Solow-Swan Model.}
 \label{sse12}
\end{figure}

The classical Solow-Swan capital accumulation model is governed by the nonlinear differential equation (\ref{sse1}). Due to its nonlinearity, this equation resists many standard analytical techniques, which motivates the use of decomposition and transform-based approaches. Setting $\frac{dk(t)}{dt} =0$ in (\ref{sse1})  identifies the equilibrium points:
$$k=0, k=\left(\frac{p}{q}\right)^{\frac{1}{1-\mu}}.$$
These equilibria are shown in Figure~\ref{sse12}. While varying the parameters $p$ and $q$ alters the scale and numerical values of the non-zero equilibrium, the qualitative behaviour of the solution remains invariant.

Figure \ref{sse12}. Graphical representation of the equilibria of the Solow-Swan model defined in Equation (\ref{sse1}). To solve Equation \eqref{sse1}, apply the ST:
$$\Scal \left[\frac{dk}{dt}\right]=p\Scal \left[k^{\mu}(t)\right]-q\Scal \left[k(t)\right].$$
Using the identity for the Caputo fractional derivative (Equation \eqref{sumud19}):
$$K(u)-k(0)=u\left(p\Scal \left[k^{\mu}(t)\right]-qK(u)\right).$$
Then, construct the variational iteration scheme as:
\begin{equation}\label{sse7}
K_{n+1}(u) = K_n(u) + \lambda(u) \left(\frac{K_n(u) - k_0}{u} - p \mathcal{S}[N[k_n]] + q \mathcal{S}[k_n]\right), \quad n \in \mathbb{N},
\end{equation}
where $N[k_n]= k_n^{\mu},$ and $\lambda(u)$ is the Lagrange multiplier. Treating the nonlinear term as restricted, variation yields
\begin{equation}\label{sse8}
\lambda(u) = -u.
\end{equation}
Substituting \eqref{sse8} into \eqref{sse7} and taking the inverse ST leads to the recursive formula:
\begin{eqnarray*}
K_{n+1}(u)&=&K_n(u)+\lambda(u)\bigg(\frac{K_n(u)-k_0}{u}-p\Scal \left[N[k_n]\right] + q\Scal \left[k_n\right]\bigg), 
\end{eqnarray*}
initialized with $k_0(t)=k_0.$ The nonlinear term is decomposed using the Adomian method. Assume the decomposition:
\begin{equation}\label{sse9}
k_n = \sum_{i=0}^{n} w_i,
\end{equation}
and apply the Adomian expansion:
\begin{equation}\label{sse10}
N[k_n] = \sum_{i=0}^{n} A_i, \quad \text{with} \quad A_i = \frac{1}{i!} \left[ \frac{d^i}{dx^i} f\left(\sum_{n=0}^{\infty} x^n w_n \right)\right]_{x=0}.
\end{equation}
The first few Adomian polynomials for $k_n^{\mu}$ are:
\begin{equation}\label{sse11}
\begin{cases}
A_0 = w_0^{\mu},\
A_1 = \mu w_1 w_0^{\mu - 1},\
A_2 = \mu w_2 w_0^{\mu - 1} + \frac{\mu(\mu - 1)}{2} w_1^2 w_0^{\mu - 2},\
\vdots
\end{cases}
\end{equation}
This yields the recursive solution:
$$\begin{cases}
w_0(t)=k(0)=k_0,\\
w_{n+1}(t)= \Scal^{-1}\left[u\left(p\Scal \left[A_n\right] - q\Scal \left[w_n\right]\right)\right],
\end{cases}$$
producing the sequence:
$$\begin{cases}
w_0&=k_0,\\
w_1&=\left(pk_0^{\mu}-qk_0\right)t,\\
w_2&=\left(pk_0^{\mu}-qk_0\right)\left(p\mu k_0^{\mu-1}-q\right)\frac{t^{2}}{2!},\\
w_3&=\left(pk_0^{\mu}-qk_0\right)\left\{\left(p\mu k_0^{\mu-1}-q\right)^2-\frac{p\mu(\mu-1)}{2}k_0^{\mu-2} \right\}\frac{t^{3}}{3!},\\
\vdots
\end{cases}$$
The solution to the Solow-Swan model is then approximated by the series:
\begin{eqnarray}\label{sse13}
k(t)&=&\displaystyle \lim_{n\rightarrow \infty}k_n=\displaystyle \lim_{n\rightarrow \infty}\sum_{n=0}^{\infty}w_n=k_0+\left(pk_0^{\mu}-qk_0\right)t+\left(pk_0^{\mu}-qk_0\right)\left(p\mu k_0^{\mu-1}-q\right)\frac{t^{2}}{2!}\\
&&+\left(pk_0^{\mu}-qk_0\right)\left\{\left(p\mu k_0^{\mu-1}-q\right)^2-\frac{p\mu(\mu-1)}{2}k_0^{\mu-2} \right\}\frac{t^{3}}{3!}+\cdots\nonumber
\end{eqnarray}

\begin{figure}
\centering
\includegraphics[width=9.0cm ,height=5.5cm]{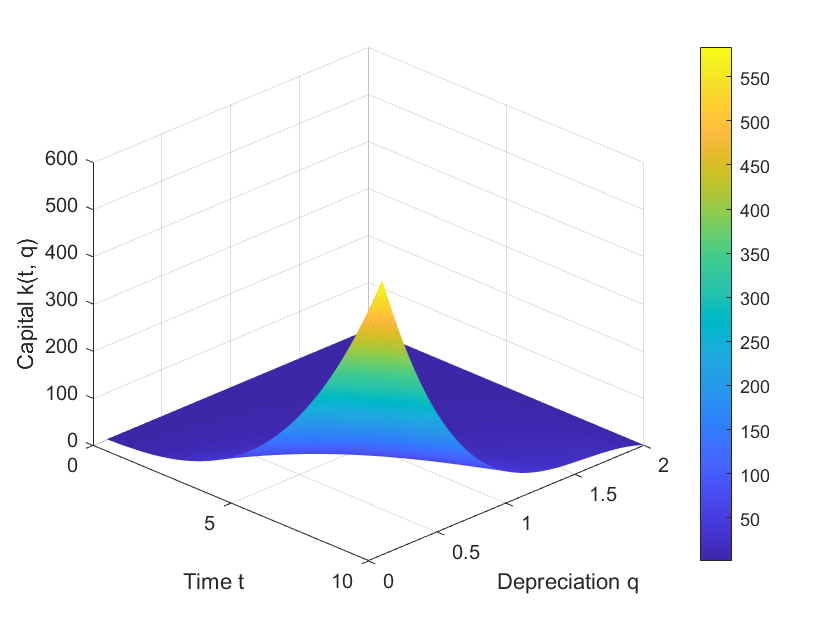}
 \caption{Graph of (\ref{sse13}) for $k(t, q)$ versus $t$ and $q$ ($p$ and $\mu$ fixed).}
 \label{sse15}
\end{figure}

\begin{figure}
\centering
\includegraphics[width=9.0cm ,height=5.5cm]{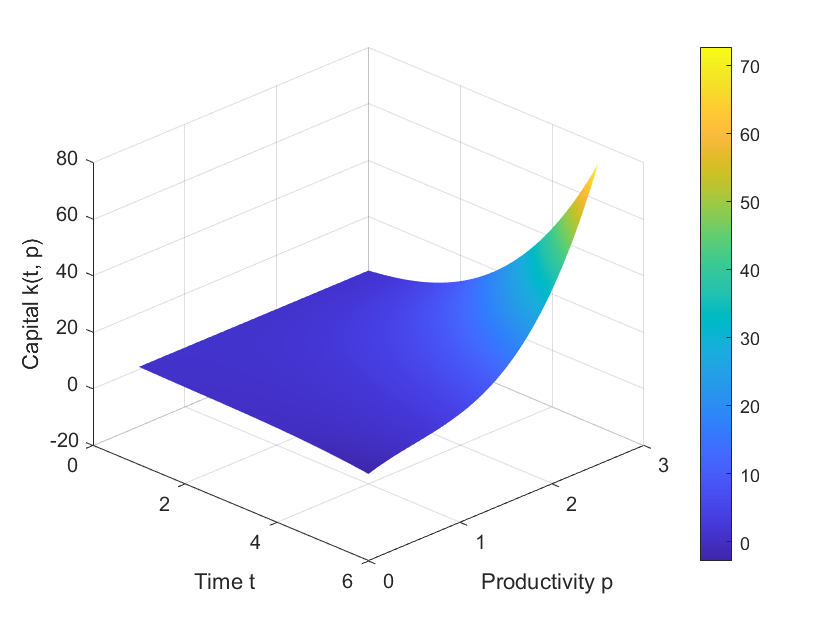}
\caption{Graph of (\ref{sse13}) for $k(t, p)$ versus $t$ and $p$ ($q$ and $\mu$ fixed).}
 \label{sse16}
\end{figure}

Using statistical data from the OECD Economic Outlook 117 \cite{OECD2025}, along with values reported in studies employing the Solow-Swan model with Cobb-Douglas-type production functions \cite{Brunner1, BarroSala2004}, parameter values were assigned for the MATLAB simulation. These reflect plausible productivity coefficients and capital elasticity parameters calibrated for developed economies. In the absence of exact empirical data, constant values within standard economic ranges were selected to illustrate the qualitative behaviour of the fractional-order model. Five terms of the series expansion are used:
$$k(t)\approx \displaystyle \sum_{n=0}^{4}w_n= w_0 + w_1 + w_2 + w_3 + w_4.$$
This truncation is chosen as a balance between computational efficiency and sufficient accuracy for capturing the qualitative behaviour of the capital dynamics. Figure \ref{sse15} illustrates the behaviour of the capital accumulation function $k(t,q)$ over time $t$ and varying values of the parameter $q,$ with fixed values of productivity $p,$ capital elasticity $\mu,$ and initial capital $k_0.$ The plot is generated from a truncated series solution of the nonlinear Solow-Swan model, capturing the first few terms of the series expansion (\ref{sse13}). Figure \ref{sse16} presents the approximate solution $k(t,p)$ to the nonlinear Solow-Swan growth model, based on a truncated Adomian series expansion. The capital stock $k(t)$ is plotted as a function of time $t$ and productivity parameter $p,$ while the depreciation rate $q$ and capital elasticity $\mu$ are held constant.

The surface plots of $k(t,p)$ and $k(t,q)$ reveal the contrasting roles of productivity and depreciation in shaping capital dynamics. As seen in the plot of $k(t,p),$ holding $q$ and $\mu$ constant, increases in productivity $p$ lead to accelerated capital accumulation over time, highlighting the positive impact of technological progress and investment efficiency on economic growth. Conversely, the plot of $k(t,q),$ with fixed $p$ and $\mu,$ shows that higher depreciation rates $q$ slow down or diminish capital growth, underscoring the adverse effects of capital obsolescence and the importance of maintenance and reinvestment. Together, the plots illustrate the dual influence of growth-enhancing and growth-retarding forces within the Solow-Swan framework, emphasizing the need for balanced policies that promote innovation while mitigating capital erosion.

\begin{figure}
\centering
\includegraphics[width=9.0cm ,height=5.5cm]{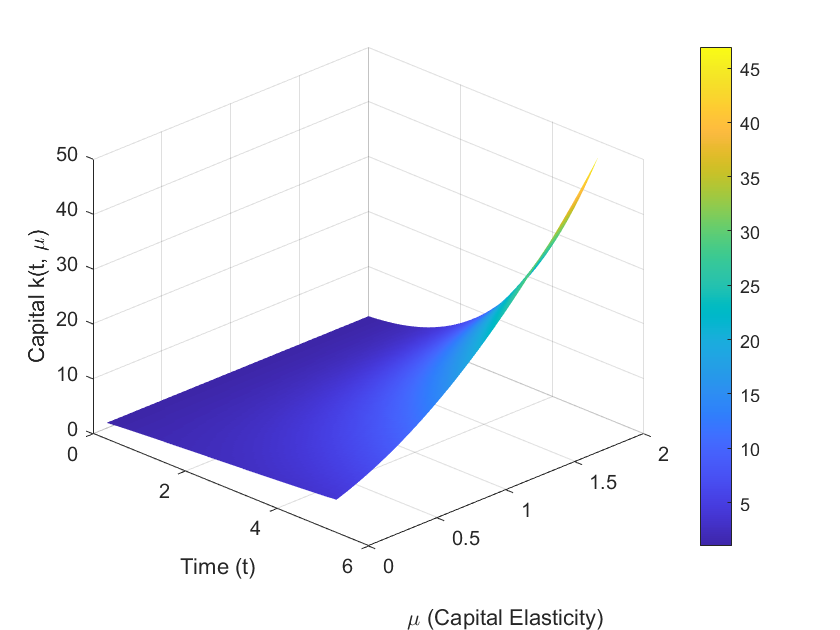}
\caption{Graph of (\ref{sse13}) for $k(t, \mu)$ versus $t$ and $\mu$ ($p$ and $q$ fixed).}
 \label{sse17}
\end{figure}

 Figure \ref{sse17} illustrates the impact of varying the capital elasticity parameter $\mu$ on capital accumulation over time, with productivity $p$ and depreciation rate $q$ fixed. The figure indicates that for small values of $\mu,$ capital grows slowly, reflecting diminishing returns to capital. As $\mu$ increases, the growth path steepens significantly, indicating that higher capital elasticity enhances the effectiveness of capital in driving output. This visual evidence underscores the critical role of $\mu$ in determining the pace and sustainability of economic growth within the Solow-Swan model framework.

\subsection{The Solow-Swan Model with Caputo Fractional Derivative}
The classical Solow-Swan model assumes that the evolution of capital over time depends solely on the current state of the system, neglecting any influence of past states. However, many real-world economic processes exhibit memory, inertia, or path dependency, which are the properties that the classical model cannot capture effectively. To incorporate these effects, we reformulate the Solow-Swan model using a Caputo fractional derivative, which generalizes the concept of differentiation to non-integer orders and inherently captures memory through its non-local structure. The Caputo fractional derivative is particularly suitable for economic modelling due to several important features. First, it preserves a key property of the classical derivative, such as the derivative of a constant is zero, making it compatible with traditional economic boundary and initial conditions. Second, it naturally accommodates initial values in the same form as ordinary differential equations, which facilitates both theoretical and numerical analyses (see, e.g., \cite{OdibatZ}). Furthermore, under certain smoothness conditions, it maintains a duality relationship with the Riemann-Liouville derivative, broadening its analytical utility.

The fractional Solow-Swan model in the sense of Caputo is given by:
\begin{equation}\label{sse2}
^C_0D^{\alpha}k(t) = k(t)\left(pk^{\mu-1}(t) - q\right), \quad k(0) = k_0,
\end{equation}
where $^C_0D^{\alpha}$ denotes the Caputo fractional derivative of order $\alpha\in(0, 1].$ When $\alpha=1,$ the model reduces to the classical integer-order Solow-Swan model, validating this extension as a true generalization.
Applying the ST to equation (\ref{sse2}) yields:
$$\Scal \left[^C_0D^{\alpha}k\right]=p\Scal \left[k^{\mu}\right]-q\Scal \left[k\right],$$
and invoking the known transform property of Caputo fractional derivative produces:
$$\frac{\Scal [k]-k_0}{u^{\alpha}}=p\Scal \left[k^{\mu}\right]-q\Scal \left[k\right].$$
The iterative correction functional is constructed by employing the variational iteration method:
\begin{equation}\label{sse3}
K_{n+1}(u) = K_n(u) + \lambda(u)\left(\frac{K_n(u) - k_0}{u^{\alpha}} - p\Scal \left[N[k_n]\right] + q\Scal \left[k_n\right]\right), \quad n \in \mathbb{N},
\end{equation}
where $N[k_n]= k_n^{\mu},$ and the Lagrange multiplier is determined as:
$$\lambda(u) = -u^{\alpha}.$$
Substituting into (\ref{sse3}) and taking the inverse ST yields:
$$k_{n+1}(t)=k_0+\Scal^{-1}\left[u^{\alpha}\left(p\Scal \left[N[k_n]\right] - q\Scal \left[k_n\right]\right)\right],$$
with initial term $k_0(t)=k_0.$ Applying the Adomian decomposition approach generates the sequence:
$$\begin{cases}
w_0&=k_0,\\
w_1&=\left(pk_0^{\mu}-qk_0\right)\frac{t^{\alpha}}{\Gamma(\alpha+1)},\\
w_2&=\left(pk_0^{\mu}-qk_0\right)\left(p\mu k_0^{\mu-1}-q\right)\frac{t^{2 \alpha}}{\Gamma(2 \alpha+1)},\\
w_3&=\left(pk_0^{\mu}-qk_0\right)\left\{\left(p\mu k_0^{\mu-1}-q\right)^2-\frac{p\mu(\mu-1)}{2}k_0^{\mu-2} \right\}\frac{t^{3\alpha}}{\Gamma(3\alpha +1)},\\
\vdots
\end{cases}$$
Hence, the analytical solution to the fractional Solow-Swan model is given by:
\begin{equation}\label{sse18}
\begin{split}
k(t) &= \lim_{n \to \infty} \sum_{n=0}^\infty w_n \\
&= k_0 + \left(pk_0^{\mu} - qk_0\right)\frac{t^{\alpha}}{\Gamma(\alpha+1)} + \left(pk_0^{\mu} - qk_0\right)\left(p\mu k_0^{\mu-1} - q\right)\frac{t^{2\alpha}}{\Gamma(2\alpha+1)} \\
&+ \left(pk_0^{\mu} - qk_0\right)\left[\left(p\mu k_0^{\mu-1} - q\right)^2 - \frac{p\mu(\mu - 1)}{2}k_0^{\mu-2} \right]\frac{t^{3\alpha}}{\Gamma(3\alpha + 1)} + \cdots
\end{split}
\end{equation}
The inclusion of the Caputo fractional derivative introduces a memory-dependent mechanism into capital accumulation, which allows past capital dynamics to influence present behaviour. This leads to a more realistic representation of economic systems, especially in scenarios where investment decisions and productivity improvements have lingering effects. Compared to the classical model, the fractional version adds greater flexibility in modelling transitional dynamics, enabling richer interpretations of convergence rates, persistence of shocks, and long-run economic trajectories. Thus, the fractional Solow-Swan model, by embedding memory via the Caputo derivative, provides a powerful framework for capturing the long-term interdependencies inherent in real-world economic growth processes.

\begin{figure}
\centering
\includegraphics[width=9.0cm ,height=5.5cm]{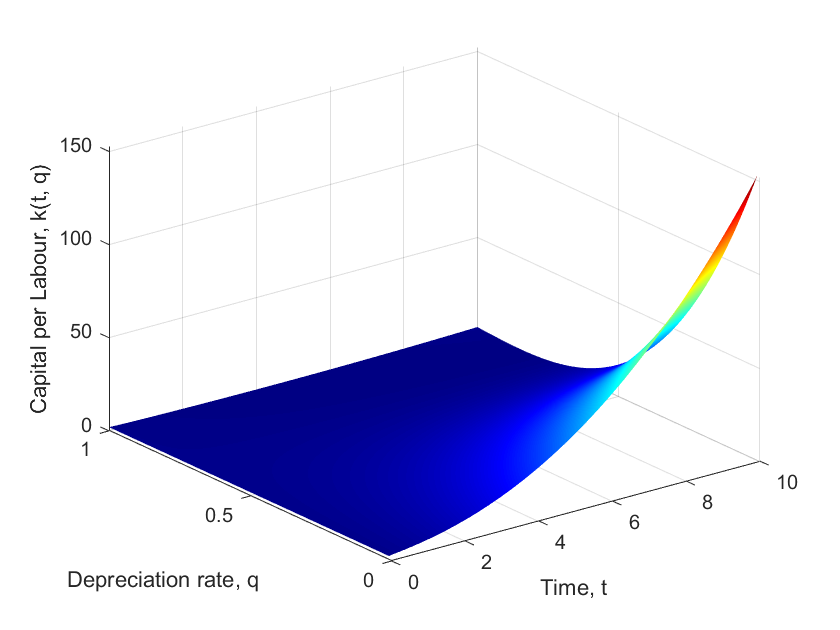}
 \caption{Graph of (\ref{sse18}) for $k(t, q)$ with constant $p, \alpha,$ and $\mu.$}
 \label{sse19}
\end{figure}

Figure \ref{sse19} illustrates the dynamic behaviour of capital accumulation over time $t$ as the depreciation rate $q$ varies, while productivity $p,$ capital elasticity $\mu,$ and the fractional order 
$\alpha$ are held constant. It indicates that for lower values of $q,$ capital per labour grows more rapidly, reflecting minimal erosion of capital stock. Conversely, as $q$ increases, the growth of $k(t)$ diminishes significantly, highlighting the dampening effect of higher depreciation. The incorporation of the fractional-order parameter $\alpha$ introduces memory into the system, leading to smoother transitions and more gradual capital dynamics compared to the classical model. This visualization underscores the sensitivity of long-run capital accumulation to depreciation and the value of fractional calculus in capturing historical dependence in economic systems.

\begin{figure}
\centering
\includegraphics[width=9.0cm ,height=5.5cm]{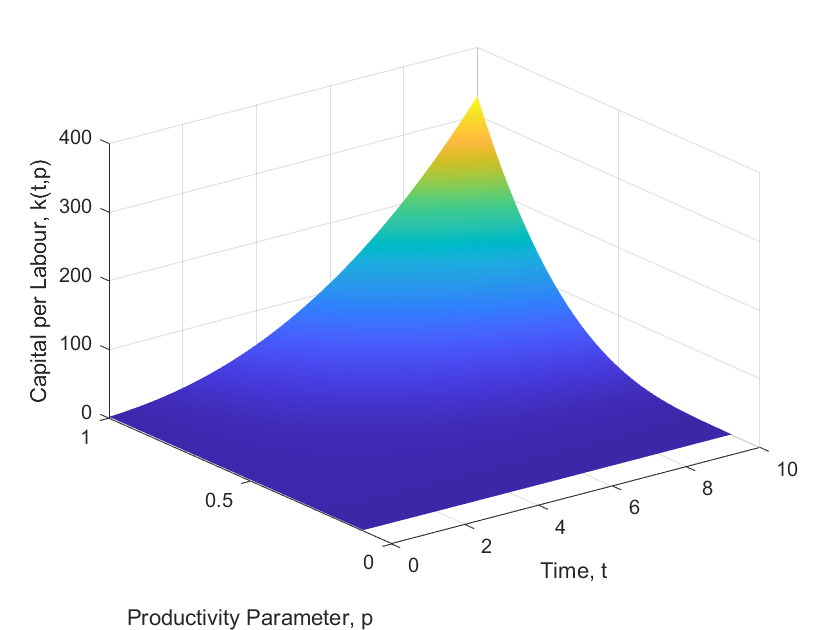}
\caption{Graph of (\ref{sse18}) for k(t, p) with constant $q,$ $\alpha,$ and $\mu.$}
 \label{sse20}
\end{figure}

Figure \ref{sse20} illustrates the evolution of capital per labour over time as the productivity parameter 
$p$ varies, while the depreciation rate $q,$ fractional order $\alpha,$ and capital elasticity $\mu$ remain constant. The figure shows that higher values of $p$ significantly enhance capital accumulation, as productivity amplifies the returns to existing capital. In contrast, lower values of $p$ result in slower or even stagnant capital growth due to insufficient productive output to offset depreciation. The use of a fractional-order derivative introduces a memory effect, reflecting the influence of historical investment decisions on current capital dynamics. This behaviour departs from the classical Solow-Swan model by capturing persistent effects and smoother transitions, making it more suitable for modelling real-world economic systems where the past has a non-negligible impact on present outcomes.

\begin{figure}
\centering
\includegraphics[width=9.0cm ,height=5.5cm]{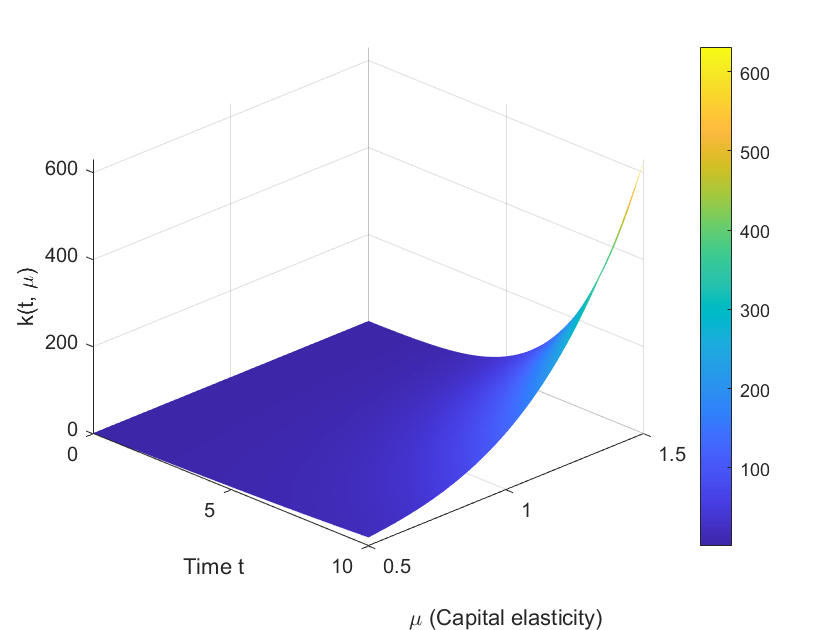}
\caption{Graph of (\ref{sse18}) for Capital Accumulation $k(t, \mu)$.}
 \label{sse22}
\end{figure}

With fixed values of productivity $p,$ depreciation rate $q,$ and fractional order $\alpha,$ Figure \ref{sse22} illustrates how capital accumulation evolves over time for varying values of the capital elasticity parameter $\mu.$ The surface exhibits a steepening pattern as $\mu$ increases, reflecting enhanced responsiveness of output to capital input in the production function. For lower values of $\mu,$ capital accumulation remains subdued over time, consistent with diminishing returns. In contrast, higher values of 
$\mu$ result in rapid capital growth, signifying a shift toward stronger returns to scale. This visualization highlights the sensitivity of long-run economic growth to the capital elasticity parameter within the memory-aware fractional framework, showing how structural characteristics of the economy influence growth trajectories under persistent memory effects.

\begin{figure}
\centering
\includegraphics[width=9.0cm ,height=5.5cm]{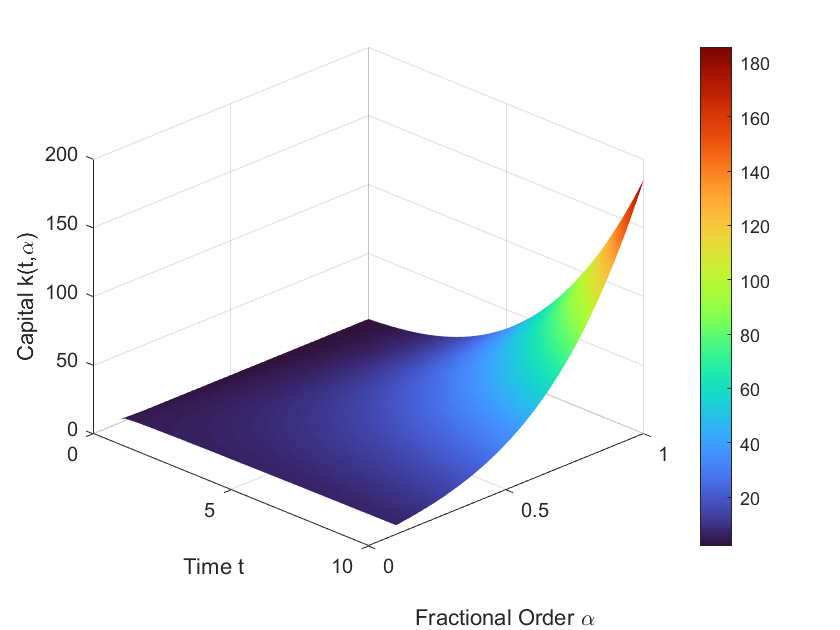}
\caption{Graph of (\ref{sse18}) for $k(t,\alpha)$ with $p, q, \mu$ constant.}
 \label{sse21}
\end{figure}

Figure \ref{sse21} demonstrates how capital accumulation evolves over time for varying values of the fractional order $\alpha,$ with the productivity parameter $p,$ depreciation rate $q,$ and capital elasticity 
$\mu$ held fixed. As $\alpha$ increases from values close to $0$ toward $1,$ the system transitions from strong memory effects to behaviour resembling the classical Solow-Swan model. For smaller $\alpha,$ capital accumulation is more gradual due to the dominance of historical inertia, reflecting longer lasting effects of past investment. As $\alpha$ approaches $1,$ the model responds more rapidly to changes in economic parameters, resembling the memoryless nature of the classical model. This fractional framework thus offers a richer, more realistic representation of economic growth dynamics by incorporating memory and path-dependence, which are especially relevant in long-term investment and capital formation processes.

\subsection{Parameter Analysis in Terms of Capital $K$ and Labour $L$}
The equilibrium points of the classical and fractional Solow-Swan models given by equations (\ref{sse1}) and (\ref{sse2}) are
$$k = 0\ \mbox{and} \ k=\left(\frac{p}{q}\right)^\frac{1}{1-\mu}.$$ 
The equilibrium point $k=0$ is unstable since any small positive perturbation $(k>0)$ leads to a positive growth rate of capital, i.e. $\frac{dk}{dt}>0$ (or $^C_0D^{\alpha}k(t)>0$ in the fractional model); causing capital to increase over time. On the other hand, the non-zero equilibrium $k=\left(\frac{p}{q}\right)^\frac{1}{1-\mu}$ is asymptotically stable. At this point, the graphs of $k(t)$ typically exhibit an inflection, and the right-hand sides of both (\ref{sse1}) and (\ref{sse2}) reach their maximum, indicating a turning point in the capital accumulation process.

In the classical Solow-Swan model, stability implies that as $t\rightarrow\infty,$ the capital-labour ratio 
$k(t)$ converges to a stable steady-state value say $k_1.$ Given that the labour force grows exponentially as
 $L(t)=L_0e^{\psi t},$  it follows that total capital $K(t)=k(t)L(t)$ will also grow asymptotically at the same exponential rate:
$$K(t)\approx k_1L(t)=k_1L_0e^{\psi t}.$$
Incorporating memory through the Caputo fractional derivative in (\ref{sse2}) modifies the dynamics of convergence. The parameter $\alpha\in(0, 1]$ introduces a hereditary effect, meaning that past states influence current growth. While the long-term equilibrium remains $k=\left(\frac{p}{q}\right)^\frac{1}{1-\mu},$ the path to convergence is more gradual and nuanced in the fractional case. Specifically, a smaller $\alpha$ implies slower convergence to the steady-state, reflecting delays and persistent effects in capital accumulation, the phenomena that are often observed in real economies but not captured by the classical model.

Thus, the fractional Solow-Swan model offers a richer framework by allowing adjustment speeds and memory effects to be explicitly modelled. It shows that when capital per worker is below its equilibrium, growth accelerates but retains a 'memory' of its earlier states. Eventually, $k(t)$ stabilizes at the equilibrium, and total capital $K(t)$ grows proportionally with labour, consistent with balanced growth paths observed in empirical growth literature.

\section{\bf Conclusion} 
This study extends the classical Solow-Swan model by incorporating memory effects through the use of fractional-order Caputo derivatives. The traditional model, formulated using ordinary differential equations, assumes that the rate of capital accumulation depends only on the current state of the economy. However, many economic processes exhibit path dependence and delayed responses, which are not captured in the classical formulation. By introducing fractional derivatives, the model accounts for such memory effects, offering a more realistic representation of capital dynamics over time.

The study analyzed the roles of key parameters, namely the production elasticity $\mu,$ scaling constants $p
$ and $q,$ and the fractional order $\alpha,$ in shaping the behaviour of the capital-labour ratio $k(t).$ Our results showed that the model admits two equilibrium points, with the non-zero equilibrium $k=(p/q)^{1/(1−\mu)}$ being asymptotically stable in both the classical and fractional cases. However, the presence of the fractional order $\alpha$ slows down the rate of convergence to this equilibrium, emphasizing the importance of memory in economic evolution. 

Through analytical expansion and $3D$ numerical simulations, the study demonstrated how varying the parameters affects capital accumulation trajectories. When $\alpha$ is fixed, the influence of $p, q,$ and 
$\mu$ aligns with expected economic intuition: higher productivity ($p$) or output elasticity ($\mu$) leads to faster growth, while higher depreciation or labour growth ($q$) suppresses capital accumulation. Conversely, varying $\alpha$ while holding other parameters constant revealed the dampening effect of memory on short-term growth.

For practical purposes, five terms of the series expansion were used in the simulation, which are sufficient to capture the dominant qualitative effects of memory and parameter interactions in the fractional Solow-Swan model. Increasing the number of terms beyond this point yields diminishing returns in accuracy and does not significantly affect the qualitative features of the solution.

In conclusion, the fractional Solow-Swan model provides a powerful generalization of the classical growth framework. It retains the essential features of long-run growth while capturing the inertia and persistence observed in real-world economies. This approach opens up new directions for empirical calibration and further theoretical exploration in macroeconomic modelling, particularly in contexts where historical dependence and gradual adjustments play a crucial role.

\vspace{1.0cm}
\footnotesize
\noindent{\bf Abbreviations}:\\
 ADM: Adomian Decomposition Method\\
 FDEs: Fractional Differential Equations\\
 ST: Sumudu Transform\\

 \vspace{1.0cm}
\noindent{\bf Declarations}\\
\noindent{\bf Competing Interests}:\\
The authors state that there are no potential conflicts of interest..\\

\noindent{\bf Funding}:\\
 There are no funding agencies.\\

\noindent{\bf Authors’ contributions}\\
MOA was responsible for the conceptualization of the study. MOA and SM contributed substantially to the manuscript drafting. KJD was responsible for proofreading the manuscript. All authors contributed to the revision of the manuscript and approved the final version.

\noindent{\bf Ethical approval}:\\
 Ethical approval was not required, as the study did not involve any human participants, animals, or industry-related data.\\

\noindent{\bf Availability of data and materials}:\\ The data used in this study are available from the author upon reasonable request.

\end{document}